# Latent Heat and Raman Spectrum Change on Phase Transition of Water Including Calcium Chloride

Dongha Kim[1*] and Changyoun Lee[1]

[1]Korean Minjok Leadership Academy, Hoengseong, Korea
*logan.dh.kim@gmail.com

**Abstract** This study investigates how Raman O–H stretching signatures may be able to track latent-heat–related energetics during water's phase transitions. Prior work on latent heat and Raman analysis of phase change is reviewed to establish theoretical approach. Building on a framework that links O–H frequency shifts to hydrogen-bond strength, we relate entropy/enthalpy variations to temperature-dependent changes in peak position and integrated intensity. Experiments on pure water and impurity-doped samples—specifically $CaCl_2$—were performed under controlled temperatures, and spectra were deconvoluted and integrated over defined bands. The Raman metrics exhibit reproducible temperature dependence that captures the transition window, while $CaCl_2$ induces systematic shifts and measurable modifications in the apparent latent-heat signature. These results highlight Raman spectroscopy as a quantitative probe of thermodynamic evolution and impurity effects in water and related systems.

**Key Words: Raman Spectroscopy, Latent Heat, Ice, Phase Transition, Calcium Chloride**

## I. INTRODUCTION

Water, one of the most abundant substances on Earth, covers approximately 71% of the planet's surface and constitutes 55–60% of human body mass. Despite its ubiquity, water continues to present unresolved scientific questions, particularly in relation to its anomalous behaviors during phase transitions. Among the tools available for investigating molecular-scale processes, Raman spectroscopy has proven highly effective in probing hydrogen-bond dynamics and vibrational modes of water molecules [1]. During the transition from solid to liquid, Raman spectra exhibit distinct modifications in both peak position and intensity [2,3]. A notable observation is the change of Raman peak intensity near the melting point of ice [3]. While spectral shifts are commonly attributed to structural rearrangements and reorganization of hydrogen-bond networks, the mechanistic basis for intensity enhancement during melting has not been fully established.

One possible explanation involves the thermodynamic concept of latent heat. Latent heat, defined as the energy absorbed or released during a phase transition at constant temperature, has been extensively measured in bulk water. However, its microscopic correlation with hydrogen–oxygen vibrational dynamics remains insufficiently explored. As illustrated in Figure 1(a), the heating curve of water shows plateaus at 0 °C and 100 °C, corresponding to latent heat absorption during melting and vaporization, respectively. In the case of melting, latent heat is invested in disrupting the ordered ice lattice and reorganizing the hydrogen-bond network [4].

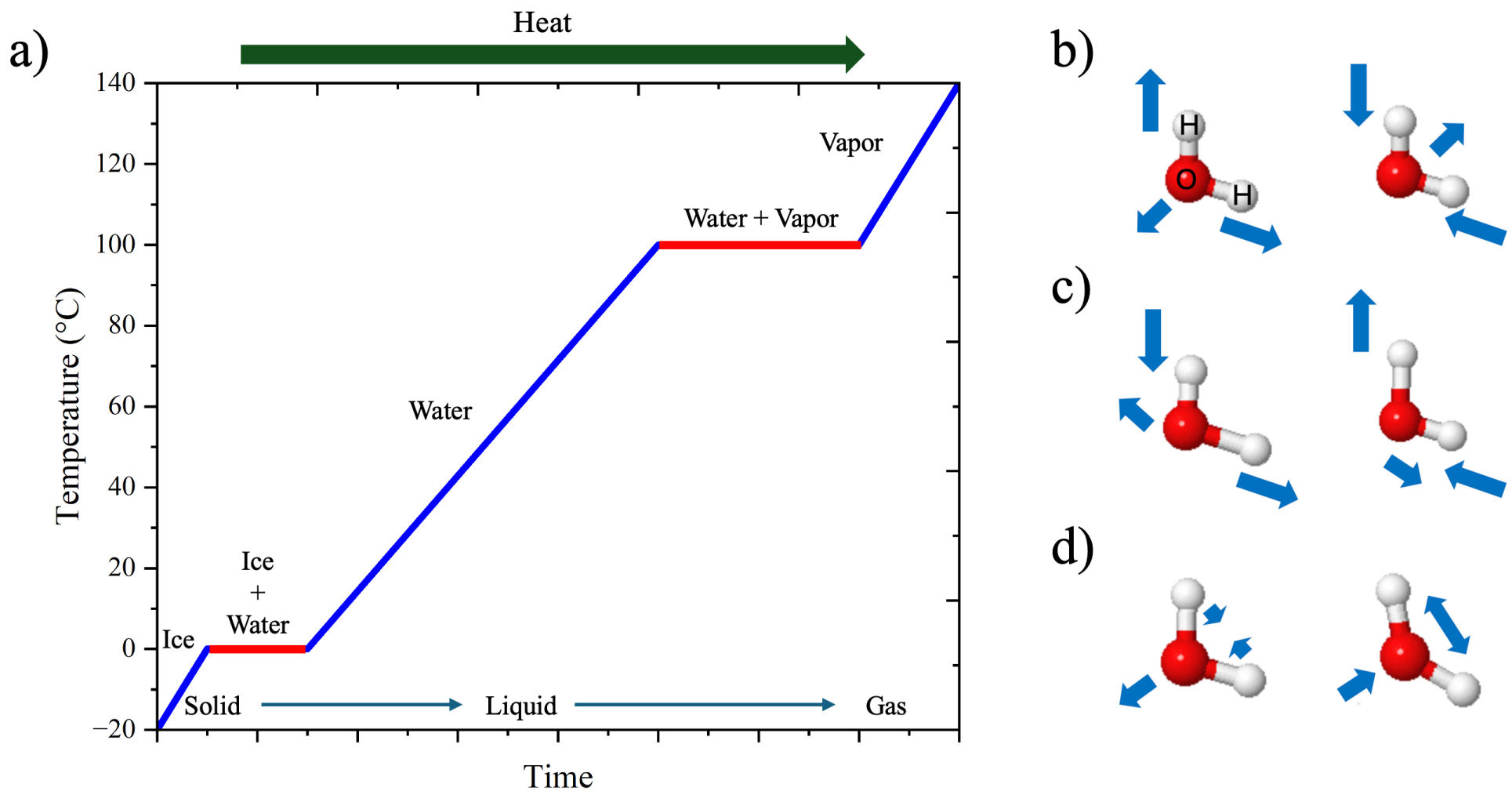

Figure 1. a) Heating curve of water, showing temperature vs. time as heat is applied. Plateaus at 0 °C and 100 °C correspond to phase transitions where latent heat is absorbed without a temperature rise: melting (ice to liquid water) and vaporization (liquid to vapor). (b–d) Schematic representations of molecular vibrational modes of $H_2O$: (b) symmetric stretching, (c) asymmetric stretching, and (d) bending mode.

This energy influx may enhance influx may contribute indirectly to vibrational excitations of O–H bonds, thereby contributing to the amplification of Raman scattering intensities. The principal vibrational modes of water molecules are the symmetric stretching, asymmetric stretching, and bending modes of the O–H bond, as shown in Figure 1(b)–(d) [5]. Alterations in hydrogen-bonding due to latent heat exchange during melting modify the contributions of these vibrational modes, which are captured in Raman measurements. While this coupling between thermodynamic energy exchange and spectroscopic response has been qualitatively discussed, a quantitative framework linking latent heat to Raman spectral variations remains lacking.

From an experimental perspective, Raman measurements of ice undergoing phase transition reveal notable spectral characteristics. Figure 2 shows the schematic diagram of the Raman spectroscopy setup used to investigate phase transitions in water. A 532 nm laser beam is directed via a reflecting mirror and focused onto the sample surface. The water sample, prepared in the ice phase, is placed on a temperature-controlled stage, while liquid nitrogen is employed to achieve and maintain low-temperature conditions across the phase-transition region. When ice samples are measured by Raman spectroscopy, as shown in Figure 3, without active temperature control of the ice, observable changes in the Raman signal occur as the ice undergoes a phase transition. These spectral variations suggest a correlation between the phase-transition-induced modification of hydrogen-bond structures and the latent heat released or absorbed during the transition.

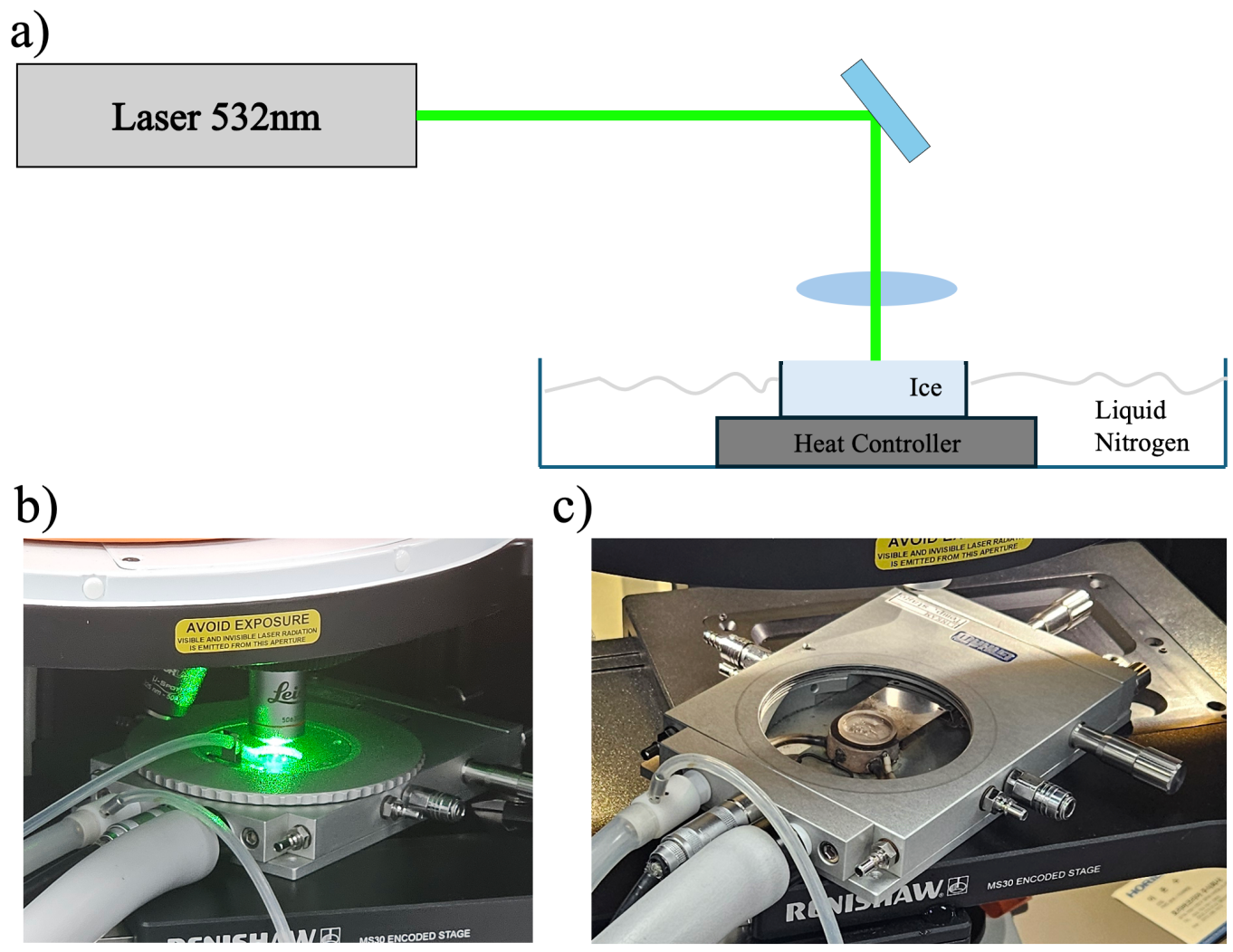

Figure 2. a) Schematics of the Raman setup(Renishaw inVia Qontor), b) and c) Pictures of the actual setup and Raman measurement. Thermostat-controlled stage was used for Raman measurement.

The objective of this study is to analyze the relationship between latent heat and Raman spectral changes during the ice–water phase transition. Particular emphasis is placed on the amplification of Raman intensity as a potential manifestation of latent heat–driven vibrational activity. Comparative experiments were performed on distilled water and calcium chloride ($CaCl_2$)-doped water, enabling evaluation of hydrogen-bond modifications under different thermodynamic conditions. The results can provide new insights into the coupling of latent heat and molecular vibrational dynamics and highlight the potential of Raman spectroscopy as a quantitative tool for monitoring phase transitions in aqueous systems.

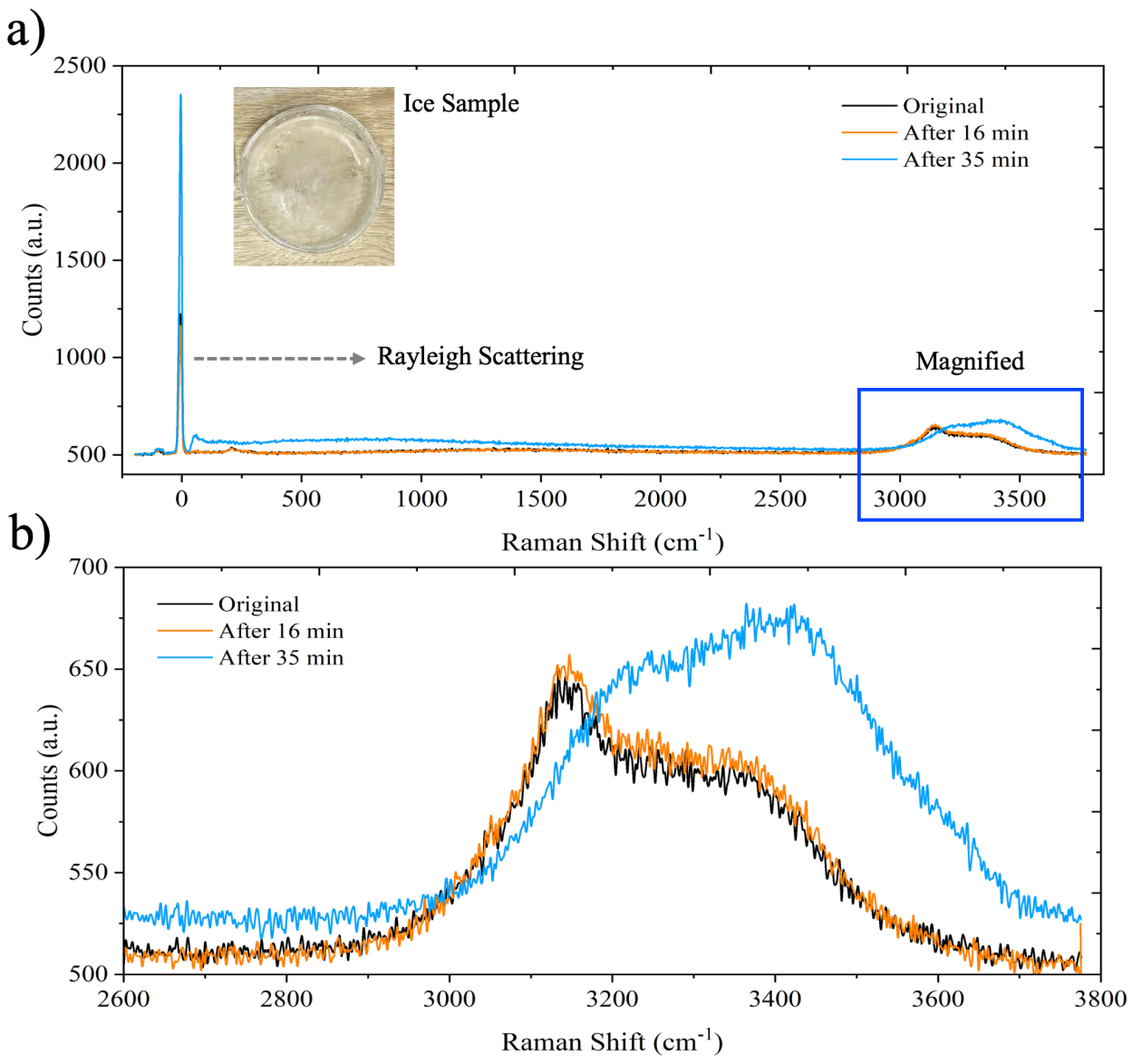

Figure 3. Raman spectra of an ice sample at different melting stages. (a) Full spectral range with Rayleigh scattering near 0 cm⁻¹ and magnified O–H stretching region (2600–3800 cm⁻¹). Inset: photograph of the ice sample. (b) Enlarged O–H stretching region showing systematic intensity decrease with melting time (0, 16, and 35 min). These spectral changes reflect hydrogen-bond network rearrangement during the ice-to-water phase transition.

## II. THEORETICAL APPROACHES ON RAMAN SPECTROSCOPY AND LATENT HEAT

Raman spectroscopy arises from the inelastic scattering of monochromatic light, where incident photons exchange energy with molecular vibrations, resulting in Stokes or anti-Stokes spectral features [6]. The intensity of Raman scattering is primarily determined by: (i) the derivative of polarizability with respect to the vibrational coordinate, and (ii) the thermal population of vibrational states, which follows Bose–Einstein statistics and is inherently temperature dependent [7]. The Stokes Raman intensity is expressed as [8]:

$$I \propto \left(\frac{\partial \alpha}{\partial q}\right)^2 \cdot \frac{1}{1-e^{-h\nu/kT}} \tag{1}$$

where α is the polarizability, $q$ is the normal vibrational coordinate, $h$ is Planck's constant, $\nu$ is the vibrational frequency, $k$ is Boltzmann's constant, and $T$ is the absolute temperature. In ice, characteristic O–H stretching vibrational modes occur at 2988, 3150, 3350, and 3506 cm$^{-1}$, reflecting the influence of hydrogen bonding [9]. These modes are sensitive to molecular rearrangements, making them suitable probes for detecting structural changes during melting. Meanwhile, the thermodynamics of Latent Heat can be described

by the general expression for heat transfer in a system:

$$Q = mc\Delta T \tag{2}$$

where $Q$ is the heat absorbed, $m$ is mass, $c$ is specific heat capacity, and $\Delta T$ is the temperature change [10]. However, during a first-order phase transition such as melting, $\Delta T = 0$, and the absorbed energy is associated not with temperature increase but with structural reorganization. Therefore, the appropriate thermodynamic relation becomes:

$$Q = mL = \Delta H = T\Delta S \tag{3}$$

where $L$ is latent heat, $\Delta H$ is enthalpy change, and $\Delta S$ is entropy change. For ice melting, the absorbed latent heat disrupts the hydrogen-bonded lattice and promotes reorganization into the liquid structure, thereby increasing entropy and enabling transient vibrational states.

The equilibrium between two phases during a transition is described by the Clausius–Clapeyron relation [10]:

$$\frac{dP}{dT} = \frac{\Delta H}{T\Delta V} \tag{4}$$

where $dP/dT$ is the slope of the coexistence line in the pressure–temperature phase diagram, $\Delta H$ is the enthalpy (in relation to latent heat) of transition, and $\Delta V$ is the volume change associated with the phase transformation. For the ice–water transition, the anomalous density difference ($\Delta V < 0$) and the finite latent heat of fusion ($\Delta H > 0$) determine the slope of the melting curve. This thermodynamic framework quantifies the energy input required to sustain phase equilibrium. The latent heat absorbed during melting reorganizes the hydrogen-bond network, thereby modifying the polarizability derivatives ($\partial\alpha/\partial q$) that control Raman intensity (Eq. 1). Substituting Eq. (3) into Eq. (4), the Clausius–Clapeyron relation provides a direct connection between macroscopic thermodynamics and microscopic vibrational dynamics:

$$\Delta H = T\Delta S = T\Delta V \cdot \frac{dP}{dT} \tag{5}$$

This enthalpy change corresponds to the latent heat that drives hydrogen-bond reorganization. An increase in $\Delta H$ can influence the population of transient vibration states, which may in turn contribute indirectly to the enhancement of Raman scattering intensity. In other words, as $\Delta H$ increases the population of transient vibrational states, the effective polarizability derivative is enhanced, producing amplification in Raman scattering intensity. Thus, by combining Eqs. (1) and (5), a theoretical framework emerges in which Raman intensity variations can be viewed as spectroscopic signatures of latent heat absorption during phase transitions. This coupling offers a quantitative pathway for predicting Raman signal amplification from fundamental thermodynamic parameters.

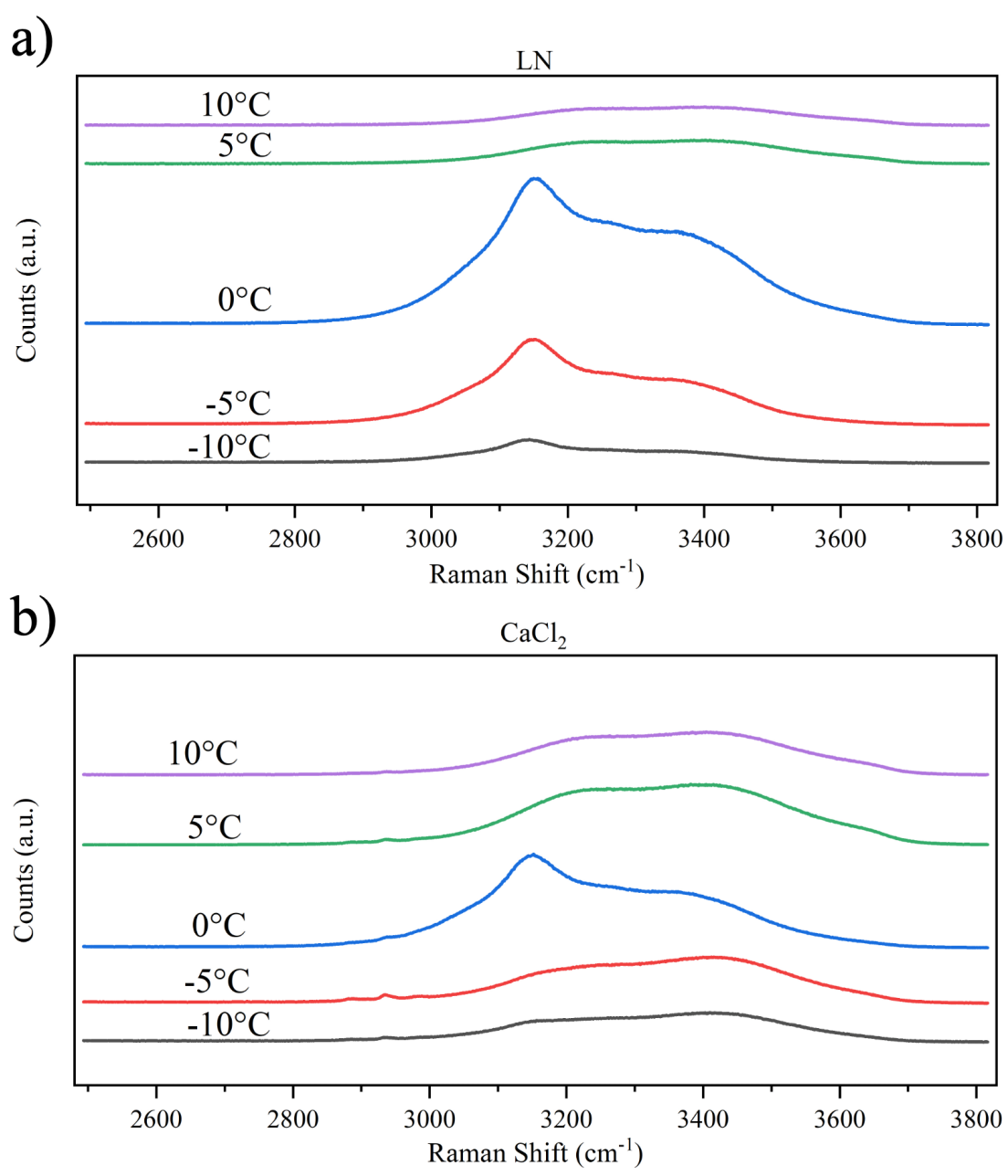

Figure 4. Temperature-dependent Raman spectra in the O–H stretching region (2600–3800 cm$^{-1}$). (a) LN-frozen ice and (b) $CaCl_2$-doped ice measured at –10 °C, –5 °C, 0 °C, 5 °C, and 10 °C. Systematic variations in peak position and intensity reveal hydrogen-bond network rearrangement and impurity effects during phase transitions.

## III. EXPERIMENTS AND MEASUREMENTS

Two types of ice samples were prepared to examine Raman spectral changes during the ice–water phase transition.

1. **Liquid nitrogen frozen ice (LN):** Deionized water was rapidly frozen by immersion in liquid nitrogen at –196 °C. This rapid quenching process minimized crystal growth and produced fine-grained ice.
2. **Calcium chloride ice ($CaCl_2$):** Ice was prepared from $CaCl_2$-saturated water to investigate the effect of ionic doping on hydrogen-bond dynamics.

Raman spectra were acquired using an in-situ Raman Spectroscope (Renishaw inVia Qontor), equipped with a 532 nm laser as the excitation source as shown Figure 2. The laser beam was focused onto the sample surface via a microscope objective. Backscattered Raman signals were collected in a 180° geometry, filtered using a notch/edge filter to remove Rayleigh scattering, and dispersed by a diffraction grating before being detected by a thermoelectrically cooled CCD camera. Samples were placed inside a sealed environmental chamber integrated with the Raman spectrometer. The chamber was equipped with both a liquid nitrogen cooling system and an embedded heater, enabling precise temperature regulation. Liquid nitrogen was continuously pumped into the chamber to achieve sub-zero temperatures, while the heater maintained controlled warming. The sample temperature was monitored and recorded at each measurement point using embedded thermocouples. For each sample type, Raman spectra were collected at five target temperatures:

–10 °C, –5 °C, 0 °C, 5 °C, and 10 °C. Measurements began at –10 °C, and the temperature was increased sequentially. At each step, the sample was equilibrated for at least 5 minutes before spectral acquisition to ensure thermal stability. Figure 4 presents representative Raman spectra of each sample at the target temperatures. Notably, the spectra at 0 °C exhibited enhanced intensity compared to those recorded at other conditions, consistent with vibrational amplification near the melting point.

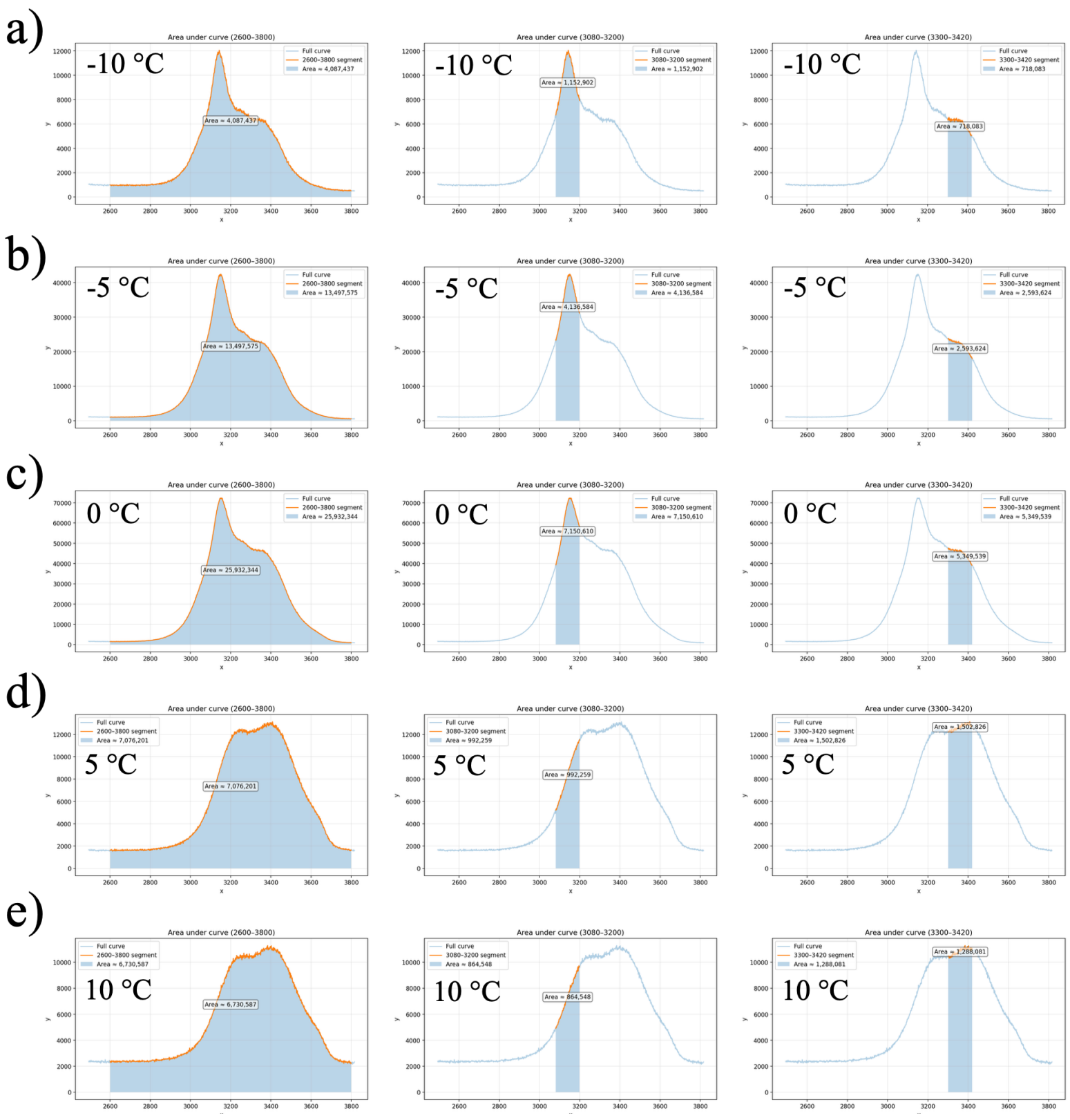

Figure 5. Raman spectra of liquid-nitrogen frozen ice in the O–H stretching region measured at –10, –5, 0, 5, and 10 °C. Left graphs display the full-range spectra (2600–3800 cm⁻¹), where the blue-shaded regions represent integrated spectral areas. The middle and right graphs highlight the integrated ranges of 3080–3200 and 3300–3420 cm⁻¹, respectively.

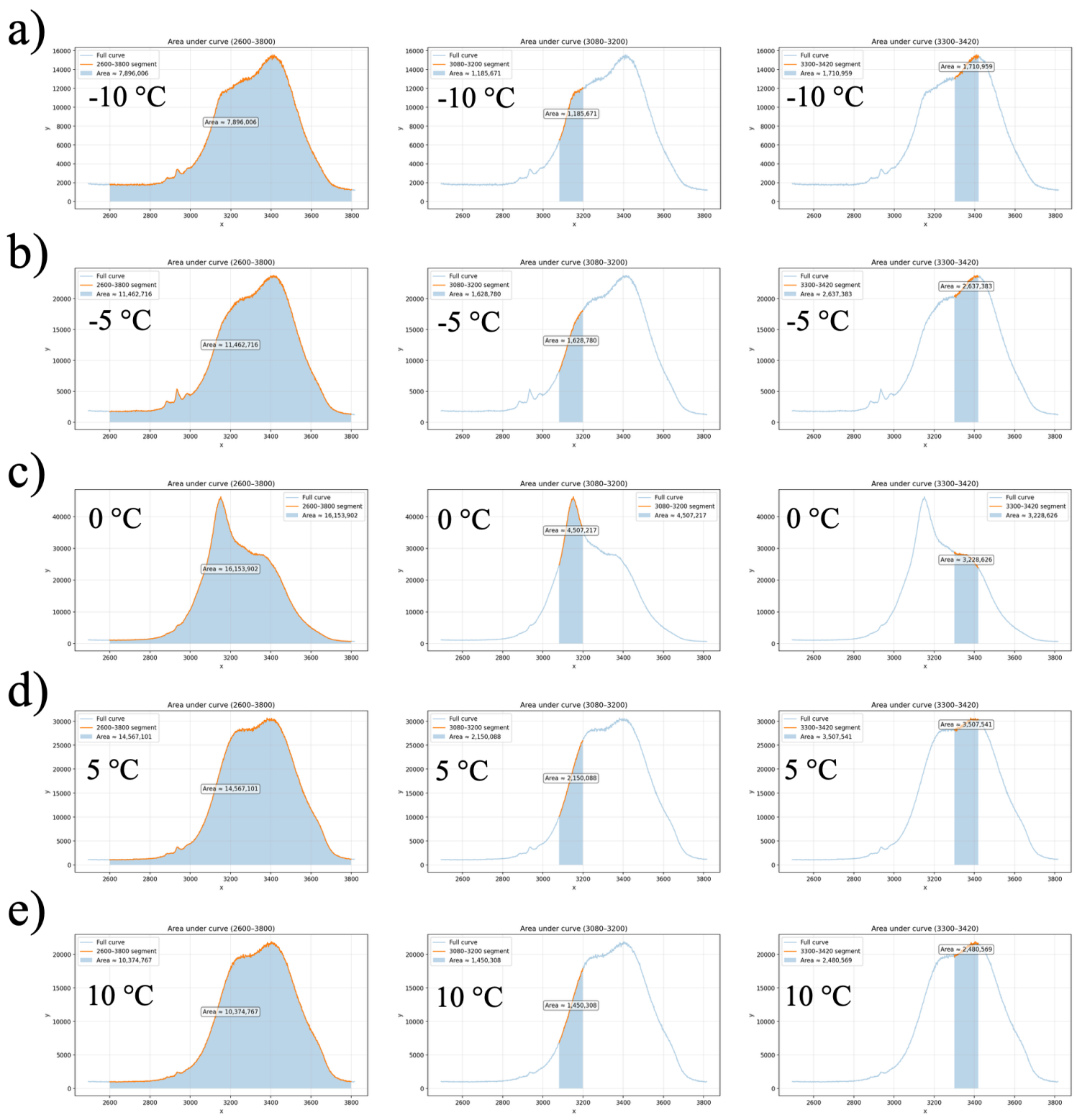

Figure 6. Raman spectra of $CaCl_2$-doped ice in the O–H stretching region measured at –10, –5, 0, 5, and 10 °C. Left graphs display the full-range spectra (2600–3800 cm$^{-1}$), with blue-shaded regions indicating integrated spectral areas. The middle and right graphs highlight the integrated ranges of 3080–3200 and 3300–3420 cm$^{-1}$, respectively.

## IV. Data Analysis

The measured Raman spectra exhibited clear enhancement near 0 °C in both sample types (LN-frozen and $CaCl_2$-doped ice) as shown in Fig. 4. However, the curvature and intensity distribution differed between the two samples. To quantify these differences, integrated Raman intensities were calculated for three spectral regions: the full O–H stretching range (2600–3800 cm$^{-1}$), the symmetric stretching region (3080–3200 cm$^{-1}$), and the asymmetric stretching region (3300–3420 cm$^{-1}$), as illustrated in Figures 5 and 6. The integrated intensities were denoted as $I_{\mathrm{sym}}$ for the symmetric mode and $I_{\mathrm{asym}}$ for the asymmetric mode. As shown in Figure 7(a)–(c), both $I_{\mathrm{sym}}$ and $I_{\mathrm{asym}}$ displayed a distinct peak near 0 °C, consistent with enhanced vibrational activity during the phase transition. Đuričković et al. proposed the use of a symmetry deviation ($S_D$) parameter to distinguish solid–liquid transitions in water based on O–H stretching features [2]. The $S_D$ parameter is defined as the ratio of integrated intensities in the symmetric and asymmetric stretching regions:

$$S_D = \frac{\Delta I_{sym}}{\Delta I_{asym}} = \int_{3080}^{3200} I(\lambda)d\lambda \,/ \int_{3300}^{3420} I(\lambda)d\lambda \tag{6}$$

where $I(\lambda)$is the Raman intensity at a given wavenumber $\lambda$. As shown in Figure 7(d) (blue), the LN-frozen ice sample exhibited a sharp change in the $S_D$ parameter near 0 °C, corresponding to the phase transition. By contrast, the $CaCl_2$-doped sample displayed a significant drop of $S_D$ value at 0 °C, suggesting that impurities alter the balance between symmetric and asymmetric O–H vibrational modes (Figure 7(d), orange). While asymmetric stretching modes typically dominate, their contribution diminished significantly in the doped sample during the transition. Both samples displayed a pronounced difference near 0 °C, followed by an increase at higher temperatures. Notably, the $CaCl_2$ sample showed a larger drop in this ratio, indicating stronger perturbation of the hydrogen-bonding environment by ionic impurities compared to pure water. This observation is significantly different from the previous reports [2].

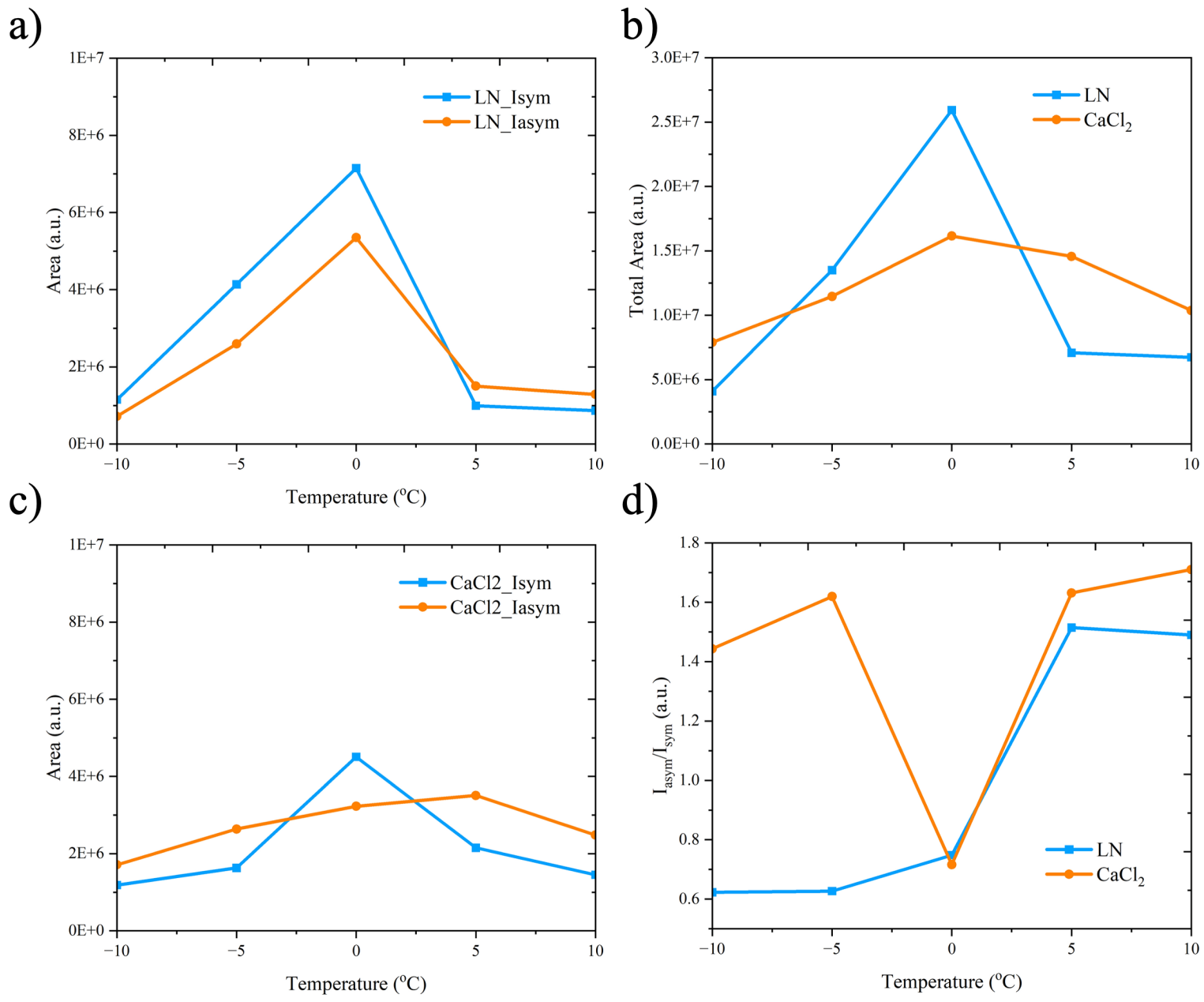

Figure 7. Temperature-dependent Raman analysis of LN-frozen and $CaCl_2$-doped ice. (a) Integrated area of the symmetric O–H stretching band ($I_{sym}$ and $I_{asym}$) for LN-frozen ice. (b) Total spectral area comparison between LN and $CaCl_2$-doped ice. (c) $I_{sym}$ and $I_{asym}$ variation for $CaCl_2$-doped samples. (d) Intensity ratio $I_{asym}/I_{sym}$ as a function of temperature for LN and $CaCl_2$ ice.

The observed Raman spectral variations can be directly linked to hydrogen-bond energetics. The relationship between hydrogen-bond energy change ($\Delta E_{\mathrm{HB}}$) and O–H stretching frequency shift ($\Delta \nu_{\mathrm{OH}}$) is:

$$\Delta E_{HB} \propto \Delta \nu_{OH} \tag{7}$$

Here $\Delta E_{HB}$ was assumed to represent the enthalpy change originating from hydrogen bonding. Then, $\Delta E_{\mathrm{HB}} \sim \Delta H = T\Delta S$ [from Eq. (3)], this yields:

$$\Delta E_{HB} \propto \Delta S \tag{6}$$

Entropy variation during the transition can thus be related to the temperature derivative of Raman intensity:

$$\frac{dS}{dT} \propto \dot{q}(T) \propto \frac{dI_{sym}}{dT} \tag{7}$$

where $\dot{q}(T)$ is heat flow rate and consequently,

$$S(T) = c_0 I_{sym}(T) + c_1 \tag{8}$$

where $c_0$ and $c_{01}$ are constant. This can be calculated as Enthalpy like metric $\Delta H(T)$ [eq. (9)].

$$\Delta H(T) = \int_{T_0}^{T} T' dS = c_1 \int_{T_0}^{T} T' dI_{sym}(T') \tag{9}$$

This $\Delta H(T)$ are proportional to latent heat (Eq. [3]). Using the measured $I_{\mathrm{sym}}$, smoothed data between –10 °C and 10 °C were processed in Python to calculate entropy changes. Figures 8(a) and (c) show the variation of $I_{\mathrm{sym}}$ for LN and $CaCl_2$ samples, while Figures 8(b) and (d) display the corresponding $dS/dT$. For LN-frozen samples, the entropy derivative exhibited a sharp divergence near 0 °C, marking the onset of the phase transition. In contrast, the $CaCl_2$-doped samples showed significant entropy fluctuations both before and after the transition, reflecting the role of impurities in broadening and perturbing the hydrogen-bond rearrangement process.

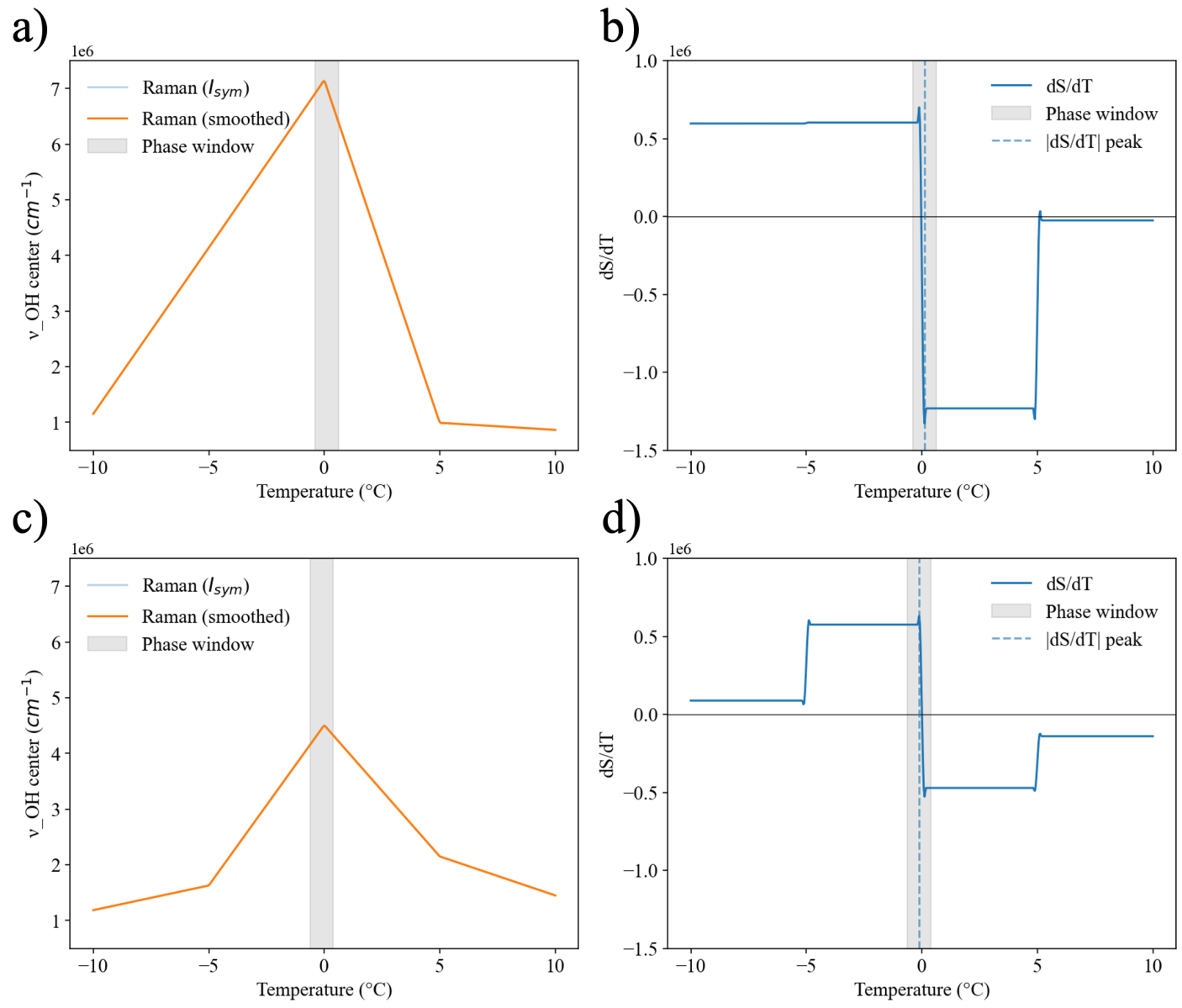

Figure 8. Thermodynamic analysis based on Raman spectral area. (a, b) LN-frozen ice and (c, d) $CaCl_2$-doped ice. Graphs (a, c) show the temperature dependence of integrated Raman intensity in the O–H stretching region (orange, smoothed line), with the shaded gray region marking the phase transition window. Graphs (b, d) present the corresponding derivatives (dS/dT), where distinct peaks highlight the phase transition signatures. The comparison illustrates the influence of $CaCl_2$ impurities on hydrogen-bond network rearrangements and entropy-related responses.

Figure 9 displays $\Delta H(T)$ based on the corresponding $dS/dT$ for both LN (a) and $CaCl_2$ (b) sample, respectively. The latent heat at the phase window (gray shaded region in Fig. 9) shows that latent heat of pure ice (LN) is higher than one of $CaCl_2$-doped ice. The Raman analysis revealed that both LN-frozen and $CaCl_2$-doped samples exhibited enhanced intensity near 0 °C, consistent with vibrational amplification during the ice–water transition. However, the spectral features and intensity ratios differed between the two samples. The $S_D$ parameter, that is, the ratio $I_{\mathrm{asym}}/I_{\mathrm{sym}}$demonstrated that $CaCl_2$ impurities significantly modified the O–H stretching modes, leading to reduced asymmetric contributions during melting. Thermodynamic interpretation linked these spectral variations to entropy changes, with the LN sample showing sharp entropy divergence at 0 °C, while the $CaCl_2$ sample exhibited broader fluctuations extending beyond the transition point. These results point towards the sensitivity of Raman intensity to latent heat–driven hydrogen-bond reorganization and emphasize the distinct influence of ionic impurities on phase transition dynamics.

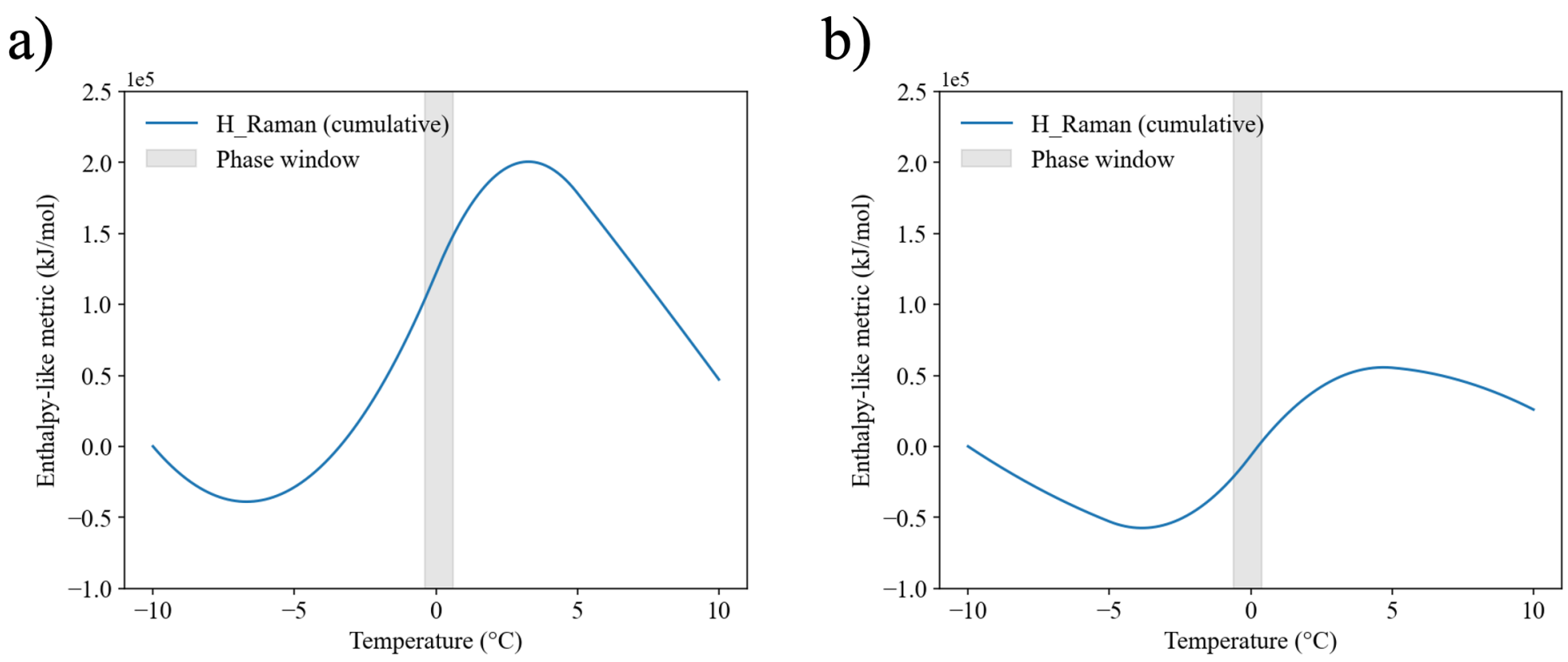

Figure 9. Enthalpy-like metric derived from cumulative Raman intensity as a function of temperature. (a) LN-frozen ice and (b) $CaCl_2$-doped ice. The shaded region marks the phase transition window. Distinct temperature-dependent trends indicate hydrogen-bond network rearrangements and impurity-induced modifications to the enthalpic response during the ice–water phase transition.

# V. Results and discussion

This study investigated the coupling between latent heat absorption and Raman spectral changes during the ice–water phase transition. The results suggest that impurities, such as calcium chloride, may reduce the latent heat required for melting compared to pure water. These variations appear to be associated with differences in entropy behavior: while both systems exhibited sharp entropy drops at the transition, the magnitude and distribution of these changes differed depending on the presence of impurities. Such distinctions could arise from impurity-induced alterations in the O-H bond network dynamics. From a macroscopic perspective, latent heat represents the total energy required to reorganize the hydrogen-bond network during a phase transition. At the microscopic level, this energy manifests as measurable shifts in Raman peak position, intensity, and integrated area, providing spectroscopic "footprints" of molecular rearrangements. By correlating Raman observables with thermodynamic quantities such as entropy and enthalpy, this work highlights Raman spectroscopy as an indirect probe of latent heat–driven structural dynamics. Further validations through employing calorimetric techniques such as differential scanning calorimetry (DSC) could provide more precise quantification of latent heat, allowing further validation of the spectroscopic–thermodynamic framework proposed here [11-12]. Overall, these observations lead to the potential of Raman spectroscopy for elucidating hydrogen-bond reorganization at the molecular scale, while leaving room for complementary methods to establish a fuller understanding of water's phase transitions.

# Acknowledgment

I am grateful to Prof. Steve Park (KAIST) for assistance with Raman spectroscopic measurements.